\documentclass[a4paper,11pt]{article}
\usepackage{graphicx}
\usepackage{amsfonts}
\usepackage{amssymb}
\begin{document}
\title{String coupling constant seems to be 1}
\author{Youngsub Yoon \\\emph{Dunsan-ro 201, Seo-gu} \\\emph{Daejeon 35242, South Korea}}
\maketitle

\begin{abstract}
We present a reasoning that the string coupling constant should be 1 from the assumption that the area spectrum derived from loop quantum gravity must be equal to the area spectrum calculated from ``stringy differential geometry.'' To this end, we will use the loop quantum gravity area spectrum constructions proposed by Brian Kong and us, and stringy differential geometry based on double field theory recently proposed by Imtak Jeon, Kanghoon Lee and Jeong-Hyuck Park.
\end{abstract}

\section {Introduction}
String coupling constant depends on the vacuum expectation value of dilaton, and has not been determined before. In this paper, we suggest that it has to be 1 from the assumption that the area spectrum derived from loop quantum gravity must be equal to the area spectrum calculated from ``stringy differential geometry.''

To this end, we will use the loop quantum gravity area spectrum constructions proposed by Brian Kong and us \cite{KongYoon}, and apply these technics to the case of string theory. However, a difficulty persists as our earlier formalism was based on 4d spacetime, and not 10d. Therefore, in our article, we will consider the case in which the degrees of freedom don't live along the compact six dimensions. At first glance, this restriction may seem to be too stringent, but string coupling constant should be the same, whether the compact six dimensions have the degrees of freedom or not. Our hope is that a rigorous derivation will come out if the formalism of Brian Kong and us is generalized to arbitrary space-time dimensions or 10d.

Also, to our purpose, ``stringy differential geometry'' recently proposed is essential \cite{projection, stringy}. The authors of these papers propose writing the low energy effective action for the closed string massless sector in a single term. This enables us to apply loop quantum gravity technics to string theory.

The paper is organized as follows. In the next section, we briefly review the loop quantum gravity area spectrum technic proposed by us and Kong. In section 3, we briefly review stringy differential geometry proposed by Jeon, Lee and Park. In section 4, we apply loop quantum gravity technic to this differential geometry and show that string coupling constant seems to be 1. In section 5, we justify our arguments in section 4. In section 6, we conclude our paper.

\section {Lightning review of the area spectrum by Kong and Yoon}
In our earlier paper \cite{KongYoon} we obtained the following formula:
\begin{equation}
\{\tilde{A}_{cd}(\vec{\tau}), \sqrt g g^{ab}(\vec{\tau'})\}=(i) \delta^{a}_{c} \delta^{b}_{d}\delta^{3}(\vec{\tau}, \vec{\tau}')\label{pseudo}
\end{equation}
where $g$ is the 3d metric and $\tilde{A}$ is the modified complex Ashtekar variable connection proposed by us and named by us the ``newer'' variable connection. The above equation was crucial to obtain the area spectrum. In fact, as $\tilde{A}$ is a linear combination of Christoffel symbol, the above formula is the direct consequence of the following formula:

\begin{equation}
\{\Gamma^{0}_{cd}(\vec{\tau}), \sqrt{g}g^{ab}(\vec{\tau'})\}=\delta^{a}_{c} \delta^{b}_{d}\delta^{3}(\vec{\tau}, \vec{\tau}')\label{Gamma}
\end{equation}
where $\Gamma$ is the Christoffel symbol. This formula will turn out to be essential to our later discussion.

\section {Stringy differential geometry}
In \cite{stringy}, Jeon, Lee and Park succeeded in writing the following well-known low energy effective action of a closed string

\begin{equation}
S=\int dx^D \sqrt{-g} e^{-2 \phi} (R+4\partial_{\mu}\phi \partial^{\mu}\phi-\frac{1}{12}H_{\lambda\mu\nu}H^{\lambda\mu\nu})
\end{equation}
into a single term as follows:

\begin{equation}
S=\int dy^{2D}e^{-2d}\mathcal{H}^{AB}S_{AB}
\end{equation}

Here, $D$ is the dimension of spacetime and $e^{-2d}$, $\mathcal{H}$ called ``generalized metric'' and $S$ are defined by the following formulas.

\begin{equation}
e^{-2d}=\sqrt{-g}e^{-2\phi}\label{e2d}
\end{equation}

\begin{equation}
\mathcal{H}_{AB} =\left(
\begin{array}{ccc}
g^{-1} & -g^{-1}B\\
B g^{-1} & g-Bg^{-1}B
\end{array}\right)\label{generalizemetric}
\end{equation}

\begin{equation}
S_{AB}=S_{BA}:=S^{C}{}_{ACB}
\end{equation}

Here, $S$ is defined by following:
\begin{equation}
S_{ABCD}:=\frac{1}{2} (\mathcal{R}_{ABCD}+\mathcal{R}_{CDAB}-\Gamma^{E}{}_{AB}\Gamma_{ECD})
\end{equation}
and indices are raised and lowered by the following constant metric:
\begin{equation}
\mathcal{J}_{AB} :=\left(
\begin{array}{ccc}
0 & 1\\
1 & 0
\end{array}\right)
\end{equation}

$\mathcal{R}$ in turn is defined by following:
\begin{equation}
\mathcal{R}_{CDAB}=\partial_A\Gamma_{BCD}-\partial_B\Gamma_{ACD}+\Gamma_{AC}{}^{E}\Gamma_{BED}-\Gamma_{BC}{}^{E}\Gamma_{AED}
\end{equation}

We will not write down here how $\Gamma$ in the above equation is defined, as interested readers can consult the original paper. The precise form is not important. The only thing that matters is that $\Gamma$ is well-defined, genuine, and covariant connection, which makes its holonomy covariant.

\section{The area spectrum}
As in \cite{KongYoon}, we can obtain a poisson bracket between $\mathcal{H}$ and $\Gamma$ by the following procedure.

\begin{equation}
\frac{\partial \mathcal{L}}{\partial(\partial^0\Gamma_{A0B})}=e^{-2d}(\mathcal{H}^{AB}+\mathcal{H}^{BA})
\end{equation}

Here, we considered the case in which $A$ and $B$ are not time indices, (i.e. not equal to 0) as in \cite{KongYoon}. The above formula implies the following:

\begin{equation}
\{\Gamma_{A0B}, e^{-2d}\mathcal{H}^{CD}\}=\delta^C_A \delta_B^D
\end{equation}

Applying (\ref{e2d}), we get:

\begin{equation}
\{\Gamma_{A0B}, e^{-2 \phi} \sqrt{-g} \mathcal{H}^{CD}\}= \delta^C_A \delta_B^D\label{theaboveequation}
\end{equation}

If we construct $\textbf{A}$, the newer variable connection out of the Christoffel symbol in the above equation, the analogous equation of (\ref{pseudo}) in ``stringy differential geometry'' case would be following:

\begin{equation}
\{\textbf{A}_{AB}, e^{-2 \phi} \sqrt{-g}\mathcal{H}^{CD}\}=(i) \delta^C_A \delta_B^D\label{finally}
\end{equation}

Now, compare the above formula with (\ref{pseudo}). If the ``newer'' variable $\textbf{A}$ and the generalized metric $\mathcal{H}$ in (\ref{finally}) (i.e. the stringy differential geometry) can be identified with the ``newer'' variable $\tilde{A}$ and the metric $g$ in (\ref{pseudo}) (i.e. the usual Riemann geometry), we can conclude that the left-hand sides of (\ref{pseudo}) and (\ref{finally}) are identified, if $e^{-2 \phi}$ is 1. Therefore, we cannot but conclude that the string coupling constant seems to be 1. This is expected as the commutation relations (\ref{pseudo}) and (\ref{finally}) determine the area spectrum, and we assumed that the area spectrums for both cases are the same. In the next section, we will talk more about these identifications between the Riemann geometry case and the stringy differential geometry case.

\section {More about the identifications}
The identification between the generalized metric $\mathcal{H}$ and the usual metric $g$ is easy to find. As the string coupling constant when $B$ is non-zero and when it is zero should be the same, we can just consider the latter case. It is easy to see that substituting $B$ by zero in (\ref{generalizemetric}) yields this identification.

The identification between the newer variable connections is more subtle. In loop quantum gravity, we consider the spin network state as the basis of Hilbert space. Basically, they are holonomies made out of the newer variable connection. It is well understood that in Riemann case the holonomy made out of the newer variable connection is covariant. (Otherwise, they would not have been the basis of Hilbert space.) Also, it is to be noted that the holonomy made out of the newer variable connection in ``stingy differential geometry'' is covariant as well, since it is well-defined gauge covariant connection. This justifies the identification.

\section {Discussion and Conclusions}
In this paper, we argued that the string coupling should be 1. We hope that further research will put our argument on more rigorous footing; our formalism must be generalized to arbitrary space-time dimensions, in particular 10d. Also, the reason why the string coupling constant doesn't run and is fixed at value 1 must be understood.
\pagebreak

\end{document}